Magneto-transport and magnetic textures in Ho/FeCoGd/β-W multilayers.

Ramesh C. Budhani[1*], Vinay Sharma[1], Ezana Negusse[1], Jacob Casey[2], Arjun K. Pathak[2], Jerzy T. Sadowski[3], and Brian Kirby[4].

1. Department of Physics, Morgan State University, Baltimore, MD -21251, USA.
2. Department of Physics, SUNY Buffalo State, Buffalo, New York, NY 14222, USA
3. Center for Functional Nanomaterials, Brookhaven National Laboratory, Upton, NY 11973, USA.
4. NIST Center for Neutron Research, Gaithersburg, MD 20878, USA.

ABSTRACT

The enhancement of interfacial Dzyaloshinskii – Moriya Interaction (DMI) in magnetic multilayers results in the stabilization of topological spin textures like chiral domain walls and skyrmions. Here we report on the evaluation of interface-driven magnetic interactions in a uniquely designed multilayer where each magnetic layer of two antiferromagnetically coupled sublattices of 3d and 4f moments is sandwiched between the layers of β-tungsten and holmium whose spin Hall angles are large but opposite in sign. The atomic and magnetic periodicity of these multilayers is established by polarized neutron reflectivity measurements and the presence of a labyrinth domain spin texture of zero remanence with x-ray photoelectron microscopy. Measurements of the Hall resistivity ($\rho_{xy}(T, H)$) together with static magnetization ($M(T,H)$) over a broad range of temperature (T) and magnetic field (H) indicate impending compensation between 3d and 4f sublattices at T > 350 K. These multilayers are characterized by a small (0.04 %) but positive magnetoresistance indicative of interface enhance scattering and a large (40 nΩ.m) and negative anomalous $\rho_{xy}(T,H)$ which results from a parallel alignment of 4f moments with the external magnetic field. No distinct scaling is seen between $\rho_{xy}(T,H)$, $\rho_{xx}(T, H)$ and $M(T,H)$ at temperatures above 200 K where the magnetization develops out-of-plane anisotropy. The field scans of $\rho_{xy}$ at T > 200 K show a distinct cusp in the vicinity of magnetic saturation. These Hall data have been analyzed in the framework of a model where a distinct topological contribution to $\rho_{xy}$ rides over the anomalous Hall resistivities of the 3d and 4f magnetic sublattices. It is suggested that this apparent topological effect results from an interfacial DMI and dominates $\rho_{xy}(T,H)$ in the temperature regime where the 3d and 4f lattices are nearly compensated.

I. INTRODUCTION

The formation, stability, and dynamics of spin textures in thin films of ferromagnetic transition metals (TM) alloyed with rare-earth (RE) elements have attracted much attention in recent years [1-6]. Although the nucleation of magnetic bubble domains in TM-RE alloys was addressed extensively half a century ago [7,8], the current interest is driven by spin-orbit-torque (SOT) physics in thin films of these alloys interfaced with a non-magnetic heavy metal. The TM-RE alloys such as Fe-Gd are ferrimagnetic (FiM), where the 3d and 4f electron spin lattices are coupled antiferromagnetically. The relative strength of magnetic interactions in these two-sublattice systems drives compensation effects where the net magnetization goes to zero, coercive field diverges, and the magnetization acquires out-of-plane anisotropy [9,10].

The weak, out-of-plane magnetization of TM-RE films in the vicinity of the compensation point is perturbed easily by magnetic torques of the spin currents produced by a strongly spin-orbit-coupled non-magnetic (SOC-NM) layer proximate to the TM-RE film [6,11-15]. The broken inversion symmetry at the TM-RE alloy – SOC-NM interface also produces a chiral exchange interaction that twists the magnetization locally to produce topologically non-trivial spin textures called skyrmions [2-4]. The dynamics of magnetic skyrmions in TM-RE thin films interfaced with 5d metals like platinum have been imaged with x-ray absorption and scattering techniques [2,3].

The current-driven dynamics of skyrmions and the twisted magnetization seen by moving electrons in the interior of a skyrmion leave their signatures in the Hall resistivity in a field regime where the magnetization is not fully saturated [16,17]. While the skyrmion Hall effect has been studied in 3d ferromagnetic multilayers [18,19] and inversion symmetry broken alloys like MnSi [20], such studies on TM-RE systems are limited [4]. This issue is important since it provides a powerful tool to electronically detect the presence and motion of skyrmion in thin-film systems.

Ferrimagnets like TM-RE alloys offer a unique advantage over other ferromagnets in controlling the nucleation threshold, size, and speed of skyrmions due to their highly suppressed net magnetization near the compensation temperature ($T_{com}$), which can be tuned by changing the alloy composition. These advantages of FiMs can be exploited effectively if individual skyrmions could be nucleated easily in these systems. In the case of 3d FMs, the nucleation threshold for topological spin textures (TST) has been optimized by enhancing the interfacial chiral exchange. For example, periodic multilayers of $(ABC)_n$ type where B is a metallic ferromagnet and A and C are SOC-NMs that produce the Dzyaloshinskii – Moriya interaction (DMI) of opposite sign host TSTs at ambient temperature [21,22]. A similar approach has been reported to enhance the interfacial DMI in ferrimagnetic films of FeCoGd by creating periodic stacks of Pt/FeCoGd/MgO [2]. However, the enhancement of interfacial DMI is partial here because only the Pt/FeCoGd interface contributes to DMI due to the large spin Hall angle of platinum and the FeCoGd/MgO interface remains inactive.

Here we use a new approach which may potentially enhance DMI in $(ABC)_n$ multilayers of ferrimagnetic TM-RE alloys. It is based on the recent finding of a large positive Hall angle ($\Theta_{SH}$) of approximately 0.14 in elemental holmium [23]. The more-than-half filled 4f shell of Ho results in the largest $\Theta_{SH}$ amongst all 4f elements. The other SOC NM of the $(ABC)_n$ stack is β-tungsten with a reported $\Theta_{SH}$ of approximately - 0.3[24]. As seen earlier in Ir/Co/Pt multilayers [25], the opposite signs of $\Theta_{SH}$ in Ho and β-W may allow for additive DMI in $(Ho/FeCoGd/β-W)_{10}$ multilayers.

## II. EXPERIMENTAL DETAILS

The β-W (1 nm)/[Ho(2 nm)/FeCoGd(4.5 nm)/β-W(2 nm)]$_{10}$/β-W (4 nm) multilayers were deposited by sputtering of Ho, W, $Fe_{75}Co_{25}$ and Gd targets in a load-lock sputtering station of base pressure of about $1 \times 10^{-6}$ Pa. The sputtering rates of $Fe_{75}Co_{25}$ and Gd were optimized to produce the $Fe_{55}Co_{19}Gd_{26}$ films by co-sputtering. All four targets remained powered during the growth of the entire stack while their shutters were opened for precalibrated time intervals. These multilayers were deposited on 10 x10 mm$^2$ and 5 x 5 mm$^2$ pieces of silicon wafers covered with 200 nm thick thermal oxide. Metal shadow masks were used to create 300 μm wide and 3000 μm long Hall bar structures for transport measurements, which were carried out down to 2 K in a 9 T physical property measurement system with a resolution of $\approx 2 \times 10^{-4}$ Ω. The static magnetization M(T, H) of the samples was measured above and below 300 K in two separate vibrating sample magnetometers. We have also used polarized neutron reflectivity measurements at ambient temperature to address the periodicity, both structural and magnetic, of these multilayers along with X-ray photoelectron microscopy to image magnetic textures.

## III. RESULTS AND DISCUSSION

Several earlier reports [4,6,9,10,15,26] of the compensation temperature of $TM_{1-x}RE_x$ alloys in thin film and bulk forms show that the $T_{com}$ increases steadily on increasing the RE concentration in the alloy. In most cases, $T_{com}$ exceeds 300 K for x > 24 %. Our recent work on thin films of FeCoGd [27] agrees with these earlier reports. Structurally, the $TM_{1-x}RE_x$ alloy films deposited by physical vapor deposition techniques are amorphous [28]. The crystallization temperature of these alloys depends on the type of RE and TM elements, and their relative concentration in the alloy [29,30]. Thicker films of TM-RE alloys deposited at a high growth rate (≥ 1.0 nm/s) also tend to show structural inhomogeneities in the form of columnar grains due to self-shadowing effects [30]. The active magnetic layer in the periodic multilayer investigated here has the stoichiometry $Fe_{55}Co_{19}Gd_{26}$ with a crystallization temperature ≥ 400 K [29,30]. Since the deposition rate for the growth of these films was kept low (< 0.15 nm/s), self-shadowing was eliminated by deposition on a rotating substrate, and the thickness of each layer is small ( 4.5 nm), we do not expect any gross structural inhomogeneities in these films. The compensation temperature of $Fe_{55}Co_{19}Gd_{26}$ is more than 350K. However, a direct comparison between the response of the single-layer FeCoGd and the multilayer would need caution because one of the spin Hall elements of the multilayer is Ho with its strong magnetic signatures. The 4f moments of metallic Ho order antiferromagnetically at approximately 130 K with signatures of spin chirality. The antiferromagnetic (AFM) phase switches to a ferromagnetic (FM) phase at T < 30 K [31,32].

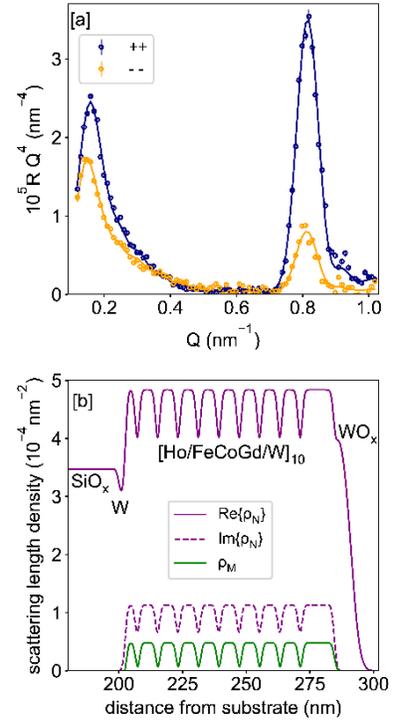

Fig. 1: (a) Spin splitting between R$^{++}$ and R$^{--}$ demonstrating sensitivity to in-plane sample magnetization parallel to external field. (b) Model fits to depth profile data for nuclear and magnetic scattering.

Polarized neutron scattering has been used extensively to address magnetic ordering in TM-RE alloys and their multilayers [33-39]. We have addressed the elemental and magnetic periodicity of the β-W (1 nm)/[Ho(2 nm)/FeCoGd(4.5 nm)/β-W(2 nm)]$_{10}$/β-W (4 nm) multilayer with polarized neutron reflectivity (PNR) measurements. Figure 1(a) shows the model-fitted reflectivity data plotted as a function of wavevector transfer Q, taken at ambient temperature in 3 T in-plane field. Data and fits are scaled by $Q^4$ to better highlight features across the entire measured Q range. PNR is used to determine the depth profiles of the nuclear scattering length density ($\rho_N$, indicative of the nuclear composition) and the magnetic scattering length density ($\rho_M$, proportional to the in-plane magnetization) of thin films and multilayers [35-37]. Non spin-flip reflectivities (++ and - -) were measured using 0.475 nm wavelength neutrons on the PBR instrument at the NIST Center for Neutron Research. Fig. 1(a) shows a large spin splitting between $R^{++}$ and $R^{--}$, demonstrating sensitivity to the in-plane sample magnetization parallel to H while the pronounced Bragg peak at Q = 0.8 nm$^{-1}$ corresponds to the average periodicity of the superlattice. Model fitting was performed using the Refl1D software package [38]. The data could not be well fit by a perfect superlattice, but it is well fit with a model that allows for significant thickness variations in the bottom-most and top-most FeCoGd layers, as shown in Fig. 1(b). The average superlattice unit cell thickness is 7.7 nm, thinner than the expected 8.5 nm. The complex nuclear profile shows $\rho_N$ close to expected values for the constituent materials. Nonzero imaginary $\rho_N$ is indicative of Gd, as it is the only element in the system with a significant imaginary component at the neutron wavelength used. The magnetic profile is shown in green, and when converted to units of magnetization, corresponds to 165 kA m$^{-1}$ for the FeCoGd layers, and a net magnetization of 136 kA m$^{-1}$ for the [Ho/FeCoGg/W]10 superlattice as a whole. It is interesting to note that the model shown corresponds to a thin (0.45 nm) and strongly negative Ho layer magnetization (-344 kA m$^{-1}$), although that strongly negative feature is smeared out by apparent interlayer roughness. However, while this profile is *consistent with* antiparallel alignment of the FeCoGd and Ho layer magnetizations, we cannot conclusively make this specific claim. Since the sensitivity here is to the depth profile averaged over a very large area in-plane, it is impossible to strictly distinguish large length scale intermixing from local interlayer roughness. As such, different choices in model parameterization could give similar or identical depth profiles that each fit the data to an indistinguishable degree but lead to different conclusions [39]. The robust conclusion that we can draw from the PNR results is that there is indeed a strongly modulated magnetic structure with FeCoGd contributing much more to the overall magnetization than does the Ho.

The magnetic domain texture of the multilayers has been imaged by means of spatially resolved x-ray magnetic circular dichroism (XMCD) at the XPEEM/LEEM endstation of the Electron Spectro-Microscopy beamline (21-ID-2) of NSLS-II. The x-ray absorption spectroscopy was performed at room temperature in a partial yield mode, with the analyzer energy window of approximately 15 eV centered on the maximum of the secondary emission peak, with the circular-left and circular-right polarization of the light, respectively, tuned to the Gd-M5 absorption edge (approximately 1173 eV). The incidence angle of the x-ray beam on the sample was about 73º to surface normal. A representative XMCD image is shown in Fig. 2. Since the contrast in the figure is related to M5 edge of Gd, its uniformity further indicates that there is no segregation of this 4f element. The labyrinthic magnetic texture with equal area coverage by domains of opposite contrast, as seen in Fig. 2, suggests an out-of-plane magnetic state of a net-zero remanence [40]. The domains in this case are known to form a manifold of metastable patterns whose topology has been studied extensively [41,42].

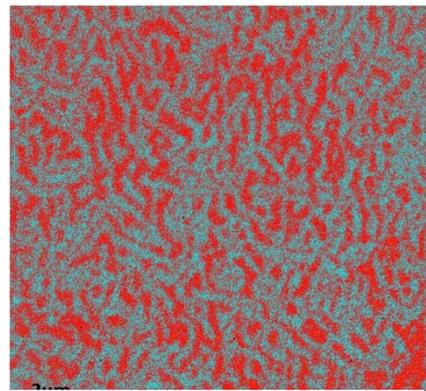

Fig. 2: X-ray magnetic circular dichroism (XMCD) image taken at the Gd-M5 absorption edge (~1173 eV) and showing stripe-like ferromagnetic domains.

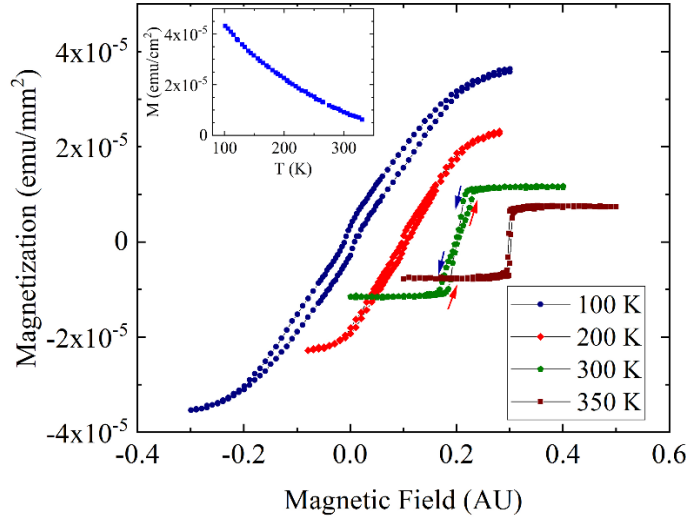

Fig.3: Out-of-plane magnetization loops of [Ho/FeCoGd/W]n multilayer measured at T = 100 K, 200 K, 300 K and 350 K. The 200 K, 300 K and 350 K loops have been given a rigid shift of 0.1T, 0.2 T and 0.3 T, respectively for clarity. Inset shows variation of saturation magnetization with temperature (1 emu/cm$^2$ = 10 A).

Our electron transport and dc magnetization studies focus on the temperature regime of 2 K ≤ T ≤ 350 K where the net magnetization of the $Fe_{55}Co_{19}Gd_{26}$ film is dominated by the contribution of 4f sublattice. The magnetic hysteresis loops of the multilayer measured at 10 K, 100 K, 300 K, and 350 K with external magnetic field oriented perpendicular to the plane of the film ($\mu_oH^\perp$) are shown in Fig. 3 where, for clarity, the data for 100 K, 300 K and 350 K have been given a rigid shift along the field axis with respect to the M(H) at 10 K. We note that the saturation field, $H_s$, decreases rapidly following the behavior of the saturation magnetization ($M_s$) as we raise the measurement temperature to 300 K. In fact, parallel field ($\mu_oH^{//}$) measurements on the multilayer revealed that at T < 300 K the magnetization has in-plane anisotropy. At 300 K however, the saturation fields for $\mu_oH^\perp$ and $\mu_oH^{//}$ configurations are identical. The M(H) loop at 300 K whose zero-remanence indicates a fine multidomain structure with an equal number of up and down magnetic domains is like that of a TM-RE alloy film that hosts bubble domains [7,8,40]. The $H_s$ drops further along with the $M_s$ on raising the temperature to 350 K. The temperature dependence of $M_s$ per unit area of the multilayer film over the temperature range of 100 to 350 K is shown in the inset of Fig. 3. The rapid drop in $M_s$ on raising the temperature is suggestive of approaching $T_{com}$.

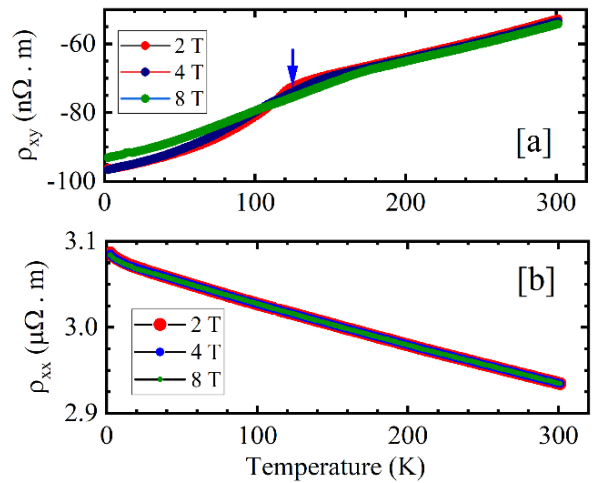

Fig. 4: (a) The Hall resistivity of the multilayer as a function of temperature from 2 to 300 K measured at 2 T, 4 T and 8 T magnetic field applied perpendicular to film plane. (b) The longitudinal resistivity as a function of temperature.

Since the Hall resistivity ($\rho_{xy}$) of a magnetically ordered material is connected to the state of its magnetization [43], we have measured the $\rho_{xy}$ of the [Ho/FeCoGd/β-W]$_{x10}$ multilayers over a broad range of temperature and strength of the external magnetic field. The measurements of $\rho_{xy}$(H) loops, which will be discussed in the subsequent section show that the Hall resistivity reaches saturation at fields ≤ 2 T at all temperatures between 2 K to 300 K. Fig. 4(a) shows the magnitude ($|\rho_{xy}|$) of the Hall resistivity at 2.0 T, 4.0 T and 8.0 T over the temperature range of 2 K to 300 K. While the $|\rho_{xy}|$ decreases on raising the temperature, the sign of $\rho_{xy}$ remains negative over the entire temperature range. Here the sign of $\rho_{xy}$ needs to be viewed in the light of the Hall resistivities of the constituent RE and TM elements, which is negative for Gd and Ho [44,45] and positive for Fe and Co [46] over the temperature range investigated here. Since the magnetization of the TM sublattice is antiparallel to the direction of the external field, the Hall voltage due to the TM sublattice will also have a negative sign. Therefore, with the data in hand, it is not possible to say which sublattice contributes the most to the value of $\rho_{xy}$. Clearly, any deviation from collinearity of two sublattices, as happens at a spin – flop transition, or due to the formation of nontrivial spin textures, will be picked up in the measurement of $\rho_{xy}$. An interesting feature of the data shown in Fig. 4(a) is the appearance of a kink in $\rho_{xy}$ (T) in the vicinity of 130 K. This feature is sharply defined in the 2 T data but becomes broader on increasing the field. While discussing the $\rho_{xy}$ (T) data of the multilayers we also need to consider the contributions of the W and Ho layers imbedded in the multilayer to the Hall coefficient. Tungsten is non-magnetic and hence its Hall coefficient is set only by the Lorentz force of magnetic field on free carriers. Since the carrier density in tungsten is high ($10^{22}/cm^3$), the Hall resistivity due to free electrons is negligible compared to the AHE of Fe and Co. However, the 4f moments of holmium order into a chiral antiferromagnetic state below 130 K through a conduction electron mediated interaction [31,32]. The AFM order changes into a FM state at T = 20 K. We believe that the kink in $\rho_{xy}$ (T) in the vicinity of 130 K is due to the AFM ordering of the Ho layers. Interestingly, the longitudinal resistivity $\rho_{xx}$(T) shown in Fig. 4(b), is insensitive to this AFM ordering. Like the resistivity of several disordered and amorphous alloys [47,48], here the temperature coefficient of $\rho_{xx}$ is negative. Further, the large value of resistivity indicates that the electron mean free path is comparable to the interatomic distance. In such a scenario, the standard Gruneisen model of electrical conduction in metals where the $\rho(T)$ is governed by electron – phonon scattering does not apply [49]. The resistivity in the limit of extreme disorder is modeled by considering the wave nature of electrons and their quantum tunneling across disordered regions in the material [50]. A tunneling scenario in the present case is facilitated by the interfaces of the multilayer and atomic scale inhomogeneities in the amorphous alloy FeCoGd. We also note that while the $\rho_{xy}$ change by about 50 % on raising the temperature from 2 K to 300 K, the corresponding change in $\rho_{xx}$ is only 1.5%. Further, Fig 4(b) shown that the out-of-plane magnetic field has virtually no effect on the $\rho_{xx}$. From these observations, it is easy to understand why the AFM order in Ho layers does not reflect in the measurements of $\rho_{xx}$. We note from Fig. 4(a) that the saturation value of $\rho_{xy}$ drops monotonically with the increasing temperature for $T > T_N$. While the out-of-plane M(T) also shows a similar behavior (Fig. 3(inset)), the slopes of the two curves are different, implying that the $\rho_{xy}$ does not scale with the magnetization.

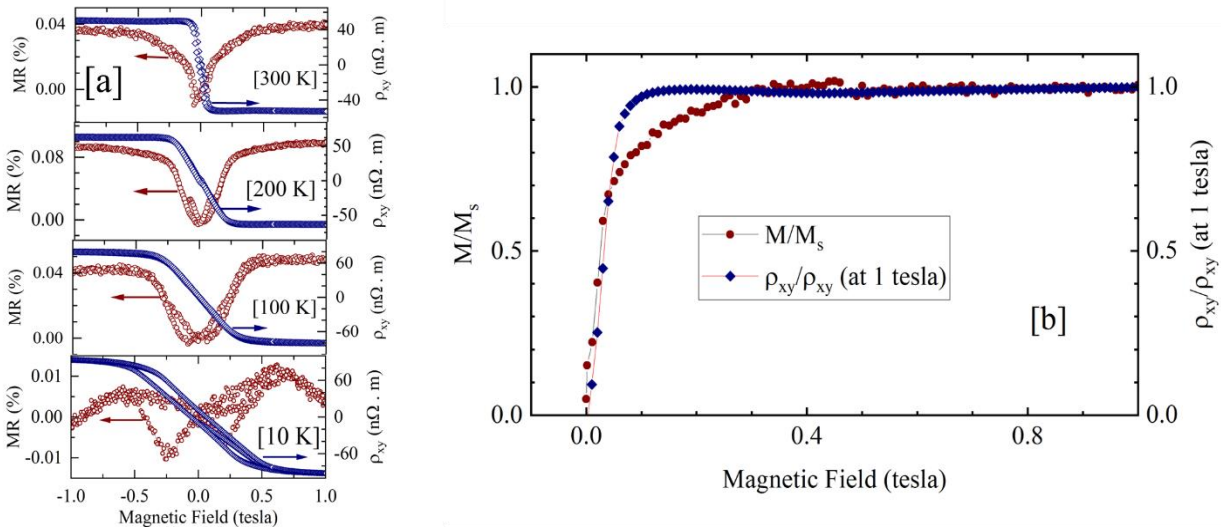

Fig. 5: (a) Loops of the out-of-plane field magnetoresistance (MR %) and anomalous Hall resistivity ($\rho_{xy}$) at 10, 100, 200 and 300 K. (b) The variations of magnetization and Hall resistivity normalized to their value at 1 T are plotted as a function of magnetic field. These measurements were performed at 300 K. The absolute value to $\rho_{xx}$ and $\rho_{xy}$ has an error of ± 10% due the uncertainty in the measurement of film thickness.

We address the possible correlation between $\rho_{xx}$ (H) and $\rho_{xy}$ (H) by measuring these resistivities at several temperatures in the field range of 0 to ±1 T. Fig. 5 shows the loops of magnetoresistance $\{(\rho_{xx}(H) - \rho_{xx}(0))/\rho_{xx}(0) \times 100\}$ and $\rho_{xy}(H)$ at 10 K, 100 K, 200 K and 300 K. As seen earlier in the temperature dependence of $\rho_{xx}$ at several fields (Fig. 4(b)), the magnetoresistance of these high resistivity amorphous alloys is very small (< 0.1%). Further, while the MR in metallic ferromagnets is negative due to suppression of spin-dependent scattering [51], the positive MR seen here is suggestive of interface dominated scattering as seen in multilayers of 3d ferromagnets [52]. At $T < 200$ K, the $\rho_{xy}(H)$ loops show distinct irreversibility at the lowest fields but the saturation fields for MR and $\rho_{xy}$ are nearly the same. We also note that at the lowest temperature (10 K), while the $\rho_{xy}$ reaches saturation, the MR starts dropping beyond the saturation field. This is perhaps due to the suppression of spin-wave excitations at the higher field [53]. At higher temperatures (> 200 K), we also notice that the way the magnetization and $\rho_{xy}$ approach saturation are different. This is clearly seen in the 300 K data plotted in Fig. 5(b). Here the $\rho_{xy}$ first surpasses the saturation value and then flattens off at high fields. The magnetization on the other hand remains well below saturation in the regime of field where $\rho_{xy}$ displays the cusp-like behavior.

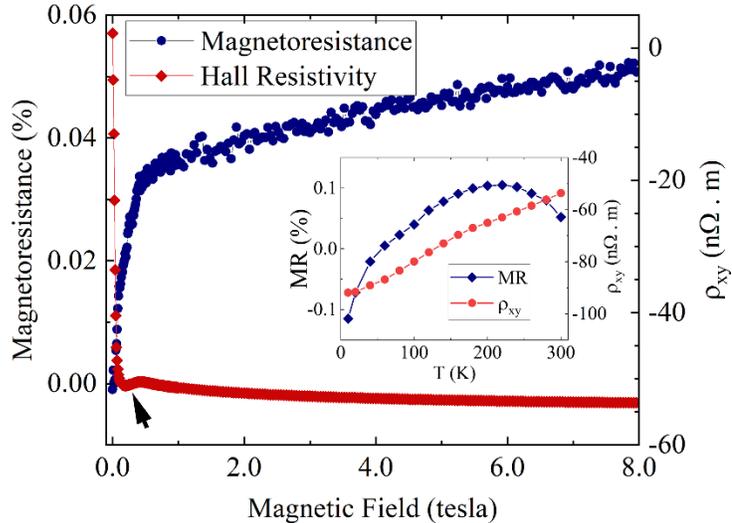

Fig. 6: The out-of-plane Magnetoresistance and anomalous Hall resistivity of the multilayer at 300 K plotted as a function of magnetic field. The black arrow at the bottom left corner of the figure highlights the cusp in $\rho_{xy}$. Inset shows variation of MR (%) and $\rho_{xy}$ at 9 Tesla as function of temperature. The absolute value to $\rho_{xx}$ and $\rho_{xy}$ has an error of ± 10% due the uncertainty in the measurement of film thickness.

We address the field dependence of MR and $\rho_{xx}$ further by measuring virgin curves in the field range of zero to 8 T. Fig. 6 shows these data taken at 300 K. The MR rises quickly from zero to 0.35% as the field is increased to 0.7 T, and beyond this field, the MR still increases but at a much slower rate (0.0025%/T). The field dependence of $\rho_{xy}$ has a distinctive cusp (indicated by a dark arrow) in the regime where the MR is approaching the state of a constant slope. To address the origin of this cusp, we first show, in the inset of Fig. 6, the variation of the saturation value of MR and $\rho_{xy}$ as a function of temperature. We note that while both MR and $\rho_{xy}$ change nearly linearly with temperature in the range of 10 to 200 K, this correlation between MR and $\rho_{xy}$ is lost at higher temperatures.

For ferromagnetic elements or alloys of a single collinear magnetic sublattice, the $\rho_{xy}$ is expressed as [43]

$$\rho_{xy} = R_0 \mu_0 H + \mu_0 R_s M_s \quad (1),$$

where $R_0$ is the ordinary Hall coefficient arising from the Lorentz force on mobile charges. In the Drude model of free carrier conduction, $R_0 = 1/ne$ where e is the electronic charge and n is the carrier density. The second term in Eq. 1 contains the effect of spontaneous magnetization on scattering and group velocity of charge carriers. The anomalous Hall coefficient $R_s$ is related to the longitudinal resistivity as $R_s = \alpha \rho_{xx} + \beta \rho_{xx}^2$, where the coefficient $\alpha$ and $\beta$ correspond to skew scattering and the effects of band structure, respectively. Further, it is also argued that the coefficient $\beta$ draws some weightage from the side-jump scattering of charge carriers [43,54]. Eq. 1 suggests that the $\rho_{xy}^{AHE}$ (= $\mu_0 R_s M_s$) should scale with the magnetization and eventually reach saturation when the system is fully magnetized. Furthermore, in the saturated state of magnetization, the $\rho_{xy}^{AHE}$ is expected to scale with the longitudinal resistivity with a power of one, or two or some intermediate value depending on the strengths of $\alpha$ and $\beta$. However, this simple relation between $\rho_{xy}$, $\rho_{xx}$, and $M_s$ of a collinear ferromagnet may not apply in the present case where two antiferromagnetically coupled magnetic sublattices compete. Moreover, the DMI at the interfaces of FeCoGd layers with Ho and $\beta$-W may make a topological contribution to the Hall resistivity as seen in the case of Pt/Co/Ir and related multilayers of 3d ferromagnets with 5d metals of large spin-orbit coupling [18,19]. We rewrite Eq.1 for the multilayer of W/FeCoGd/Ho as

$$\rho_{xy} = R_0 \mu_0 H + \mu_0 R_s^{TM} M_s^{TM} + \mu_0 R_s^{RE} M_s^{RE} + \rho_{xy}^{THE} \quad\text{---}\ (2),$$

where the second and third terms respectively are the contributions of the TM and RE sublattices to anomalous Hall. The last term in Eq. 2 is the Hall resistivity arising from the DMI driven spin textures.

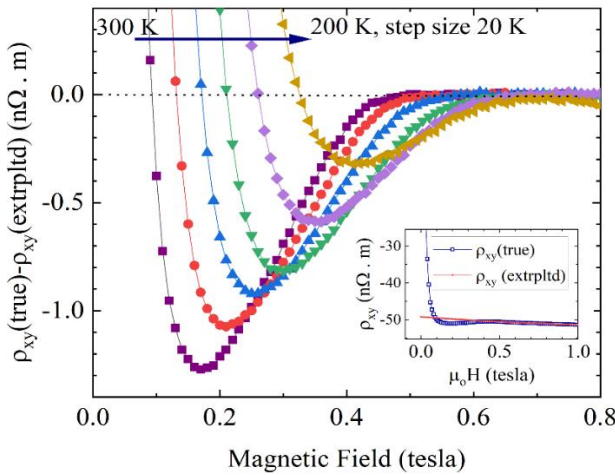

Fig. 7: The evolution of the topological Hall resistivity with magnetic field at several temperatures from 200 to 300 K. Inset shows the $\rho_{xy}$ data at 300 K over a limited range of field to highlight the procedure used to extract the data shown in the main figure.

Based on our measurements and the data reported in the literature we conclude that the sign of the anomalous Hall coefficient of Gd is negative [27,44,45] and for FeCo it is positive [46,55]. The negative sign of $R_s^{RE}$ in the temperature range investigated here (2 to 350 K) then indicates that the magnetization of the RE sublattice is parallel to the external field. The oppositely oriented TM sublattice also makes a negative contribution to $R_s$. From this argument, it is also clear that any deviations from the collinearity of two sublattices would always reduce $\rho_{xy}$ as maximum in the magnitude of $\rho_{xy}$ occurs only when the TM and RE sublattices are antiparallel. However, as clear from the $\rho_{xy}$ (H) data of Fig. 6, there is a robust excess $\rho_{xy}$ which appears as a cusp just before the saturation field is reached. We could, in principle, extract this excess Hall resistivity by noting that the first term in Eq. 2 is linear in the field and the second two terms are constant above magnetic saturation due to the negligible (≤ 0.0025 %/Tesla) magnetoresistance above $H_s$. For a single sublattice, the AHE below $H_s$ has been calculated from the scaling relation between $\rho_{xy}$ and $\rho_{xx}$ [43,56,57]. It can be expressed as

$$\rho_{xy}^{AHE} = [\alpha \rho_{xx} + \beta \rho_{xx}^2]M, \quad (3)$$

where $\alpha$ and $\beta$ are scaling coefficients as defined earlier [27,43]. However, the extraction of anomalous Hall in the field range 0 to $H_s$ is rather challenging for a two-sublattice system. Moreover, for this AFM-coupled two-sublattice system no scaling between $\rho_{xy}$ and $\rho_{xx}$ above $H_s$ is seen at T > 200 K where the cups in $\rho_{xy}$ appear. Furthermore, the correct analysis of the data in terms of Eq. 3 would need two such equations involving $\rho_{xy}$, $\rho_{xx}$ and M of each sublattice. Here we use a simple extrapolation approach to estimate the excess $\rho_{xy}$. In the inset of Fig. 7, the field-linear $\rho_{xy}$ above $H_s$ has been extrapolated to zero field. We extract the excess $\rho_{xy}$ by subtracting the measured $\rho_{xy}$ from its extrapolated value. The results of this extraction for the data taken from 200 K to 300 K are shown in the main panel of Fig.7. As we note in the figure, the excess $\rho_{xy}$ increases at higher temperatures where the system is approaching magnetic compensation.

The excess Hall resistivity below magnetic saturation in these multilayers is likely to have a topological origin. The XMCD image of Fig. 2 shows a labyrinth domain pattern at zero field and room temperature. The literature on thin films of TM-RE alloys reveals that such domains convert into bubble-like excitations on approaching magnetic saturation [7,8]. The opposite signs of the spin Hall angle in β-W and Ho layers, which periodically sandwich the FeCoGd layers are expected to induce an additive DMI which stabilizes topological spin excitations like skyrmions [21,22]. There is ample evidence of skyrmion-like spin textures in TM-RE alloys multilayered with non-magnetic heavy metals like Pt and Ta [2,3]. The excess Hall resistivity that we see here may arise as the conduction electrons traversing the magnetic textures hosting spin chirality acquire a Berry phase which creates a fictitious magnetic field. The Lorentz force experience by the conduction electrons under this magnetic field of Berry phase origin would make a distinct contribution to $\rho_{xy}$.

The DMI driven spin chirality has been addressed earlier in Pt/Gd$_{44}$Co$_{56}$/TaO$_2$ heterostructures [3] and Pt/ Gd$_{25}$Fe$_{65.6}$Co$_{9.4}$ /MgO multilayers [2]. In the former case, skyrmion of diameter ~ 10 nm are nucleated by a charge current of high density ($10^7$ A/cm$^2$) as it flows down the narrow channels of the trilayer. In the multilayer [2], skyrmions of a diameter of 180 nm are seen close to magnetic saturation where the spiral domain textures present in the fully demagnetized state disappear completely. Although to the best of our knowledge there is no Hall data available for DMI stabilized skyrmions in RE-TM alloys, we believe the robust $\rho_{xy}^{THE}$ seen here is the result of the additive DMI produced by two heavy metals of opposite Hall angle.

It is important to point out here that skyrmions also make an additional contribution to the Hall voltage as they move under the action of a large transport current. However, the current density required for depinning skyrmions is much larger ($10^7$ A/m$^2$) compared to the $J_c$ used in our experiments.

## IV. SUMMARY

We have addressed the magnetic state, robustness of atomic periodicity, and magneto-transport in a magnetic multilayer where each constituent magnetic layer consists of antiferromagnetically coupled 3d and 4f spin sublattices. This elementary magnetic unit of the multilayer is sandwiched between thin films of β-tungsten and holmium which possess spin Hall angles of opposite polarity that may promote the stabilization of topological spin textures. The zero-field ambient-temperature XPEEM images of magnetic domains reveal a sub-micron scale filamentary texture of opposite polarity indicating the zero-remanence state of an out-of-plane magnetic anisotropy film. Although the 4f spins in bulk holmium are not ordered at ambient temperature, the PNR data suggest an antiferromagnetic coupling between the 4f moments of the holmium layer and the net magnetization of the elementary magnetic unit. The measurements of longitudinal and Hall resistivities together with static magnetization over a wide range of temperature and magnetic field strength reveal some interesting aspects of the competition between 3d and 4f sublattices which are accentuated by the multilayer structure. The prominent feature of this competition is a cusp in $\rho_{xy}(H)$ which appears just below magnetic saturation and grows in strength on approaching the compensation temperature of the 3d-4f magnetic unit. A simple analysis of the anomalous Hall resistivity of our two-sublattice magnetic films where each sublattice has opposite Hall angle together with a topological contribution, which may potentially result from interfacial DMI sheds light on the origin of this cusp. Direct measurements of the strength of DMI with techniques like Brillouin light scattering would be highly beneficial in this context.

-------------------------------------------------------------------------------------------------

*Ramesh.budhani@morgan.edu

Acknowledgements

This research at Morgan State is supported by the Air Force Office of Scientific Research, Grant # FA9550-19-1-0082. We also used resources of the Center for Functional Nanomaterials and the National Synchrotron Light Source II, which are U.S. Department of Energy (DOE) Office of Science facilities at Brookhaven National Laboratory, under Contract No. DE-SC0012704. The PNR work reported here is supported by the NIST Center for Neutron Research. The work at Buffalo State was supported by the Office of Undergraduate Research Program and the faculty startup fund from the Dean's Office, School of Arts and Sciences, SUNY Buffalo State.

[3] L. Caretta, M. Mann, F. Büttner, K. Ueda, B. Pfau, C. M. Günther, P. Hessing, A. Churikova, C. Klose, M. Schneider, D. Engel, C. Marcus, D. Bono, K. Bagschik, S. Eisebitt, and G. S. D. Beach, *Fast Current-Driven Domain Walls and Small Skyrmions in a Compensated Ferrimagnet*, Nat. Nanotechnol. **13**, 1154 (2018).

[4] H. Wu, F. Groß, B. Dai, D. Lujan, S. A. Razavi, P. Zhang, Y. Liu, K. Sobotkiewich, J. Förster, M. Weigand, G. Schütz, X. Li, J. Gräfe, and K. L. Wang, *Ferrimagnetic Skyrmions in Topological Insulator/Ferrimagnet Heterostructures*, Adv. Mater. **32**, 2003380 (2020).

[5] Y. Quessab, J.-W. Xu, C. T. Ma, W. Zhou, G. A. Riley, J. M. Shaw, H. T. Nembach, S. J. Poon, and A. D. Kent, *Tuning Interfacial Dzyaloshinskii-Moriya Interactions in Thin Amorphous Ferrimagnetic Alloys*, Sci. Rep. **10**, 7447 (2020).

[6] J. A. González, J. P. Andrés, and R. López Antón, *Applied Trends in Magnetic Rare Earth/Transition Metal Alloys and Multilayers*, Sensors **21**, (2021).

[7] P. Chaudhari, J. J. Cuomo, and R. J. Gambino, *Amorphous Metallic Films for Bubble Domain Applications*, IBM J. Res. Dev. **17**, 66 (1973).

[8] R. W. Shaw, D. E. Hill, R. M. Sandfort, and J. W. Moody, *Determination of Magnetic Bubble Film Parameters from Strip Domain Measurements*, J. Appl. Phys. **44**, 2346 (1973).

[9] P. Hansen, C. Clausen, G. Much, M. Rosenkranz, and K. Witter, *Magnetic and Magneto-optical Properties of Rare-earth Transition-metal Alloys Containing Gd, Tb, Fe, Co*, J. Appl. Phys. **66**, 756 (1989).

[10] J. M. D. Coey, *Amorphous Magnetic Order*, J. Appl. Phys. **49**, 1646 (1978).

[11] K. Ueda, C.-F. Pai, A. J. Tan, M. Mann, and G. S. D. Beach, *Effect of Rare Earth Metal on the Spin-Orbit Torque in Magnetic Heterostructures*, Appl. Phys. Lett. **108**, 232405 (2016).

[12] R. Mishra, J. Yu, X. Qiu, M. Motapothula, T. Venkatesan, and H. Yang, *Anomalous Current-Induced Spin Torques in Ferrimagnets near Compensation*, Phys. Rev. Lett. **118**, 167201 (2017).

[13] N. Roschewsky, C.-H. Lambert, and S. Salahuddin, *Spin-Orbit Torque Switching of Ultralarge-Thickness Ferrimagnetic GdFeCo*, Phys. Rev. B **96**, 64406 (2017).

[14] A. V Svalov, V. O. Vas'kovskiy, I. Orue, and G. V Kurlyandskaya, *Tailoring of Switching Field in GdCo-Based Spin Valves by Inserting Co Layer*, J. Magn. Magn. Mater. **441**, 795 (2017).

[15] H. Wu, Y. Xu, P. Deng, Q. Pan, S. A. Razavi, K. Wong, L. Huang, B. Dai, Q. Shao, G. Yu, X. Han, J.-C. Rojas-Sánchez, S. Mangin, and K. L. Wang, *Spin-Orbit Torque Switching of a Nearly Compensated Ferrimagnet by Topological Surface States*, Adv. Mater. **31**, 1901681 (2019).

[16] N. Nagaosa and Y. Tokura, *Topological Properties and Dynamics of Magnetic Skyrmions*, Nat. Nanotechnol. **8**, 899 (2013).

[17] W. Jiang, G. Chen, K. Liu, J. Zang, S. G. E. te Velthuis, and A. Hoffmann, *Skyrmions in Magnetic Multilayers*, Phys. Rep. **704**, 1 (2017).
11